\documentclass[fleqn,usenatbib]{mnras}
\usepackage{newtxtext,newtxmath}
\usepackage[normalem]{ulem}
\usepackage[T1]{fontenc}
\usepackage{ae,aecompl}
\usepackage[normalem]{ulem}
\usepackage{color}

\usepackage{graphicx}	
\usepackage{hyperref}
\hypersetup{
  colorlinks   = true, 
  urlcolor     = blue, 
  linkcolor    = blue, 
  citecolor   = blue 
}
\usepackage{xcolor}

\usepackage{soul}

\def\msun{M$_{\odot}$}
\def\bg{BG\,Tri}
\def\apj{ApJ}
\def\aj{AJ}
\def\mnras{MNRAS}
\def\pasp{PASP}
\def\aap{A\&A}
\def\apjs{ApJS}
\def\msun{M$_{\sun}$}
\def\mdot{$\dot M$}

\def\grad{$^\circ$}

\title[BG Tri: a low inclination RW Sex-type nova-like]{BG Tri an example of a low inclination RW Sex-type novalike}

\author[Hern\'andez, Tovmassian, Zharikov  et al.]{M.~S. Hern\'andez $^{1}$\thanks{e--mail: 
	mercedes.hernandez@postgrado.uv.cl },  G. Tovmassian$^{2}$, S. Zharikov$^{2,3}$, B.~T. G\"ansicke$^{4,5}$, D. Steeghs$^{6,7}$, \newauthor
	A. Aungwerojwit$^{8}$, P. Rodr{\'i}guez-Gil$^{9,10}$.\\ \\
$^{1}$Instituto de F\'{i}sica y Astronom\'{i}a, Facultad de Ciencias, Universidad de Valpara\'{i}so, Av. Gran Breta\~{n}a 1111 Valpara\'{i}so, Chile\\
$^{2}$ Instituto de Astronom\'{\i}a, Universidad Nacional Aut\'onoma de M\'exico, Ensenada, Baja California, C.P. 22830, Mexico\\
$^{3}$Al-Farabi Kazakh National University, Al-Farabi Ave., 71, 050040, Almaty, Kazakhstan\\
$^{4}$University of Warwick, Department of Physics, Gibbet Hill Road, Coventry, CV4 7AL, United Kingdom.\\
$^{5}$Centre for Exoplanets and Habitability, University of Warwick, Coventry CV4 7AL, UK.\\
$^{6}$Department of Physics, Astronomy and Astrophysics group, University of Warwick, CV4 7AL, Coventry, UK.\\
$^{7}$Harvard-Smithsonian Center for Astrophysics, 60 Garden Street, Cambridge, MA 02138, USA.\\
$^{8}$Department of Physics, Faculty of Science, Naresuan University, Phitsanulok, 65000, Thailand.\\
$^{9}$Instituto de Astrof{\'i}sica de Canarias, V{\'i}a  L{\'a}ctea s/n, E-38205, La Laguna, Tenerife, Spain.\\
$^{10}$Departamento de Astrof{\'i}sica, Universidad de La Laguna, E-38206, La Laguna, Tenerife, Spain.\\
}

\date{Accepted 2021 January 28. Received 2021 January 27; in original form 2020 October 16}

\pubyear{2019}

\begin{document}

\date{Accepted 2021 January 28. Received 2021 January 27; in original form 2020 October 16}

\pagerange{\pageref{firstpage}--\pageref{lastpage}} \pubyear{2018}

\maketitle

\label{firstpage}

\begin{abstract}
We analyse a wealth of optical spectroscopic and photometric 
observations 
of the bright ($V=11.9$) cataclysmic variable \bg. 
The {\sl Gaia} DR2 parallax gives a distance $d=334(8)$~pc to the source, 
making the object one of the intrinsically brightest nova-like variables seen under a low orbital inclination angle. 
Time-resolved spectroscopic observations revealed  the orbital period of $P_{\rm{orb}}=3\fh8028(24)$. 
Its spectroscopic characteristics  resemble RW\,Sex and similar nova-like variables. We disentangled 
the H$\alpha$ emission line into two components, and show that one component forms on the irradiated face of the secondary star. We suggest that the other one originates at a disc outflow area adjacent to the L$_3$ point. 
\end{abstract}



 
\section{Introduction}
\label{intro}

Cataclysmic variables (CVs) are close binary systems comprised of a white dwarf (WD) and a low-mass star losing matter in a  Roche-lobe
overflow regime, usually creating an accretion disc around the accreting WD \citep{1995CAS....28.....W}. CVs show diverse observational
characteristics depending on fundamental physical properties, including their orbital period,  mass transfer rate; and strength of the
magnetic field of the WD. Their diversity is also in part due to the viewing angle \citep{2018AJ....156..198H}.  In non- or weakly-magnetic
systems, the accretion flow takes place in a fully-developed accretion disc.  At high mass transfer rates ($\ge 10^{-9}$~\msun\,yr$^{-1}$),
these are hot, steady-state discs  \citep{1996MNRAS.282...99B}. Therefore,  the disc thermal instability that triggers dwarf nova outbursts 
is prevented \citep{1986ApJ...305..261S,1992ApJ...394..268S,2001NewAR..45..449L}. These high mass transfer rate systems are known as 
nova-like variables (NLs), because it was initially argued that they might potentially exhibit or have undergone undetected nova eruptions
\citep{1934Obs....57..157V}. No known NL has ever been seen to erupt as a nova.  However, some NLs may resemble  nova when the nova returns to a quiescent state after the eruption \citep{1995CAS....28.....W}. They constitute a small fraction of the entire CV
population \citep[$\sim$15\ percent  in][]{2003yCat.5113....0R, 2003A&A...404..301R},
which might be result of an observational bias. The vast majority of NLs have orbital periods  above the so-called "period gap"
\citep{1983ApJ...275..713R,1998MNRAS.298L..29K,2016MNRAS.457.3867Z,2020MNRAS.492L..40A}. We exclude  CVs with moderately or highly magnetic WDs, which historically were also accounted for as NLs.  

There is a visible differentiation between NLs and  dwarf novae in terms of their orbital periods and colours \citep{2020MNRAS.492L..40A}. 
It is noticeable that NLs dominate the orbital period range 3-4h, where very few dwarf novae are observed \citep{2011ApJS..194...28K}.
However, in absolute numbers, the amount and the distribution of NLs and dwarf novae right above the period gap are comparable
\citep{2011ApJS..194...28K}. There is no good understanding why the two populations of NL and dwarf novae overlap in some period ranges, 
but not in others. NLs have accretion rates that are higher than "normal" CVs both at longer and shorter periods. They are intrinsically
very bright, and their WDs  tend to be hotter \citep{2009ApJ...693.1007T}. 

\begin{table}
  \setlength{\tabcolsep}{0.1em}

    \caption{Log of  spectroscopic observations}
    \label{tab:speclog}
    \begin{tabular}{lccccc}
       \hline \hline
      
Date   & JD$^*$ & Range~~~   & No. of   & ~~~Exp. time & ~~~Comments \\
      &         & (\AA)~~~~~ & Spectra  & (s)          &             \\ \hline
2002 Aug. 21 & 2507     &    3341-7546    &            2  & 200  &  INT       \\
2002 Aug. 23 & 2510    &    3341-7546    &            24 & 340  &  INT       \\
2002 Aug. 25 & 2512    &    3341-7546    &  15 & 120  &  INT       \\
2002 Aug. 31 & 2518    &    3341-7546    &            8 & 120  &  INT       \\
2002 Sep. 02 & 2520    &    3341-7546    &            4 & 120  &  INT       \\
2002 Sep. 03 & 2521    &    3341-7546    &            6 & 120  &  INT       \\
2003 Oct 19  & 2932         &  3784-9063      &             2  &   60   & ISIS B\&R  \\
2003 Dec. 13   & 2987    &    4257-8318    &             4 &   120   &  CA         \\
2003 Dec. 14   & 2988    &    4257-8318    &             9 &   60   &  CA         \\
2003 Dec. 15   & 2989    &    4257-8318    &             2 &   60   &  CA         \\
2003 Dec. 16   & 2990    &    4257-8318    &             10 &   60   &  CA         \\
2003 Dec. 17   & 2991    &    4257-8318    &             10 &   60, 120   &  CA         \\
2003 Dec. 17   & 2991    &    3775-6840    &             7 &   120   &  NOT         \\
2003 Dec. 23   & 2997    &    4257-8318    &             6 &   60   &  CA         \\
2003 Dec. 24   & 2998    &    4257-8318    &             6 &   60   &  CA         \\
2003 Dec. 25   & 2999    &    4257-8318    &             2 &   60   &  CA         \\
2003 Dec. 26   & 3000    &    4257-8318    &             4 &   60   &  CA         \\
2003 Dec. 27   & 3001    &    4257-8318    &             2 &   60   &  CA         \\
2004 Aug. 09   & 3226    &    4257-8318    &             2 &  180    &  CA         \\ 
2004 Aug. 10  & 3227    &    4257-8318    &             2 &  180   &  CA         \\ 
2004 Aug. 11  & 3228    &    4257-8318    &             2 &  180    &  CA         \\ 
2004 Aug. 12  & 3229    &    4257-8318    &             2 &  180    &  CA         \\ 
2004 Oct. 21  & 3300    &    4257-8318    &             2 &  120    &  CA         \\ 
2004 Oct. 23  & 3302    &    4257-8318    &             2 &  120    &  CA         \\ 
2004 Oct. 24  & 3303    &    4257-8318    &             2 &  120    &  CA         \\ 
2004 Oct. 26  & 3305    &    4257-8318    &             2 &  120    &  CA         \\ 
2004 Sep. 10  & 3623    &    4257-8318    &             2 &  120    &  CA         \\
2004 Nov. 04  & 3314    &   3775-6840     &                  2 &  120    &  NOT      \\ 
2005 Jan. 01   & 3372    &    3784-9063    &             2  &  200    & ISIS B\&R  \\
2005 Jan. 04   & 3375    &    3784-9063    &             2  &  200    & ISIS B\&R  \\
2005 Jan. 05   & 3376    &    3784-9063    &             2  &  200    & ISIS B\&R  \\
2005 Jan. 06   & 3377    &    3784-9063    &             2  &  200    & ISIS B\&R  \\
2005 Jan. 07   & 3378    &    3784-9063    &             2  &  200    & ISIS B\&R  \\
2005 Sep. 10  & 3623         &    4257-8318    &             2 &   120   &  CA         \\  
\hline
2017 Oct. 27     & ~~~~~8053~~~~~    & 3600--7300 & 3 & 1200 & Echelle \\
2017 Oct. 28     & 8054     & 3600--7300 & 3 & 1200 & Echelle \\
2017 Oct. 29     & 8055     & 3600--7300 & 3 & 1200 & Echelle \\
2018 Jan. 11     & 8129     & 3600--7300 & 13 & 1200 & Echelle \\
2018 Jan. 12     & 8130     & 3600--7300 & 12 & 1200 & Echelle \\
2018 Jan. 13     & 8131     & 3600--7300 & 3 & 1200 & Echelle \\
2018 Jan. 14     & 8132     & 3600--7300 & 10 & 1200 & Echelle  \\ \hline
       
 \end{tabular}

\begin{tabular}{l}
$^*$ 2450000+ 
JD is given at beginning of the observing night.
\end{tabular}
\end{table}

The spectra of some NL variables display persistent broad Balmer 
absorption lines, indicative of optically thick discs. However, NLs themselves come in different flavours.  
A fraction of NLs, known as  VY\,Scl stars, show  occasional states of low mass transfer rates, i.e. they become significantly 
fainter for prolonged periods of time \citep[months to years,][]{1974Obs....94..116W,2020MNRAS.494..425R}.   
However, the physical cause of these low mass transfer states is still uncertain \citep{1994ApJ...427..956L,1998ApJ...499..348K,2018A&A...617A..16S}. 

SW\,Sex stars form a class of NL variables with distinctive spectroscopic behaviour \citep{1991AJ....102..272T}. 
They mostly cluster in the 3-4\,h orbital period range \citep{10.1111/j.1365-2966.2007.11743.x}. 
\citet{1996MNRAS.282...99B}, \citet{2013MNRAS.428.3559D}, and \citet{2014AJ....147...68T} proposed an extended hot 
spot as the predominant source of emission lines  from the optically and physically thick disc. However, such interpretation 
is challenged by \citet{2007MNRAS.377.1747R,2015MNRAS.452..146R}.
A search for non-eclipsing SW\,Sex in the 3-4\,h period range revealed systems with 
two-component emission lines \citep{10.1111/j.1365-2966.2006.11245.x}, but they were inconclusive whether these are low-inclination 
SW\,Sex objects. 
Conversely,  two-component emission lines recently have been observed,  in a couple of  the UX\,UMa-type NLs (which are 
the primary concern of this paper).
For example, \citet{2017MNRAS.470.1960H}, based on high-resolution spectroscopy,  demonstrated  that NLs RW\,Sex and 
RXS\,J064434.5+334451  show at least two components in the profiles of  the Balmer emission lines. 
The narrow component with a low radial-velocity amplitude originates 
from the irradiated surface of the secondary facing the disc. 
The wide component 
is formed in an extended, low-velocity region in the outskirts of the opposite side of the accretion disc, with respect
to the collision point of the accretion stream and the disc. Recently, \citet{Subebekova20} claim, that this property is 
observed in RW~Tri.  They  compiled  a current list of similar NLs  with orbital periods 
$\geq$4 hours. At least four of them, the components of H$\alpha$ emission   
 closely resemble those seen in RW\,Sex.

Other interpretations of line provenance (including absorption features frequently flanking emission lines) in such NL systems have 
been put forward. Notably, the disc overflow model  by \citet{1996ApJ...471..949H} and disc wind interpretations and models
\citep[e.g.][]{1996AJ....111.2422P,1996Natur.382..789M} are worth mentioning. 
 
\bg\ is a bright object reported by \citet{2004AJ....127.2436W,2008PZP.....8....4K} to show an irregular, low-amplitude variability and
tentatively identified as a CV, citing ROTSE1, TYC2 and {\it ROSAT} detections. Accordingly, the object is also catalogued as 
TYC\,2298\,01538 and 1RXS J014448.4+323320. \citet{2017RMxAA..53..439M}  confirms its NL identification based on a distance estimate 
by {\it Gaia}\footnote{ $d= 334.013 \pm 7.65 $\,pc according to the revised estimate \citet{Bailer18}} \citep{2018AA...616A...1G} in combination with Galaxy Evolution Explorer {\it (GALEX) UV} magnitudes \citep{2003SPIE.4854..336M}. However, \bg\ has not been 
studied in detail until now. We report the results of the spectroscopic study of the object on the backdrop of the long term 
photometry collected by the Catalina Real-time Transient Survey and the All Sky Automated Surveys (ASAS).

\section{Observations and Reduction}
\label{sec:observations}
We  present an extensive set of multi-wavelength observations of \bg\  obtained  by us as well as a variety of data collected in surveys. 

\subsection{Photometry}
 
\begin{figure}
\setlength{\unitlength}{1mm}
\resizebox{15.cm}{!}{
\begin{picture}(130,72)(0,0)
\put(-5,0) {\includegraphics[width=8.0cm, bb=0 180 580 700, angle=0, clip]{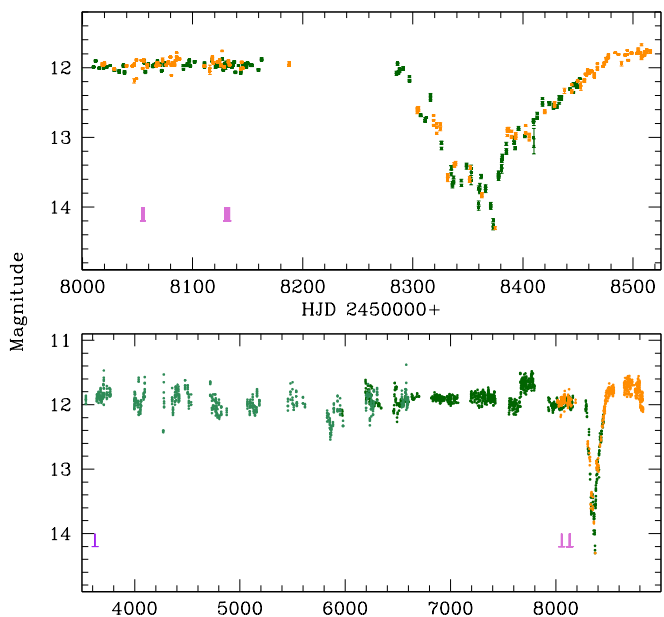}}
\end{picture}}
\caption{
 Long-term light curve of \bg\ obtained by the sky  patrol ASAS SN and CRTS surveys. The dark green and orange points are ASAS 
 data in $V$ and $g$ filters respectively, the light green  points are CRTS $V$-band data. In the lower panel, the entire data set is
 presented, while the upper shows expanded segment of the same during and around the low state. The magenta markers indicate moments of
 spectroscopic observations at the bottom of the light curve. }
\label{fig:lccurve}
\end{figure}

\begin{figure*}
\setlength{\unitlength}{1mm}
\resizebox{15.cm}{!}{
\begin{picture}(130,80)(0,0)
\put(-15,0) {\includegraphics[width=15.75cm, bb=0 160 580 520, angle=0, clip]{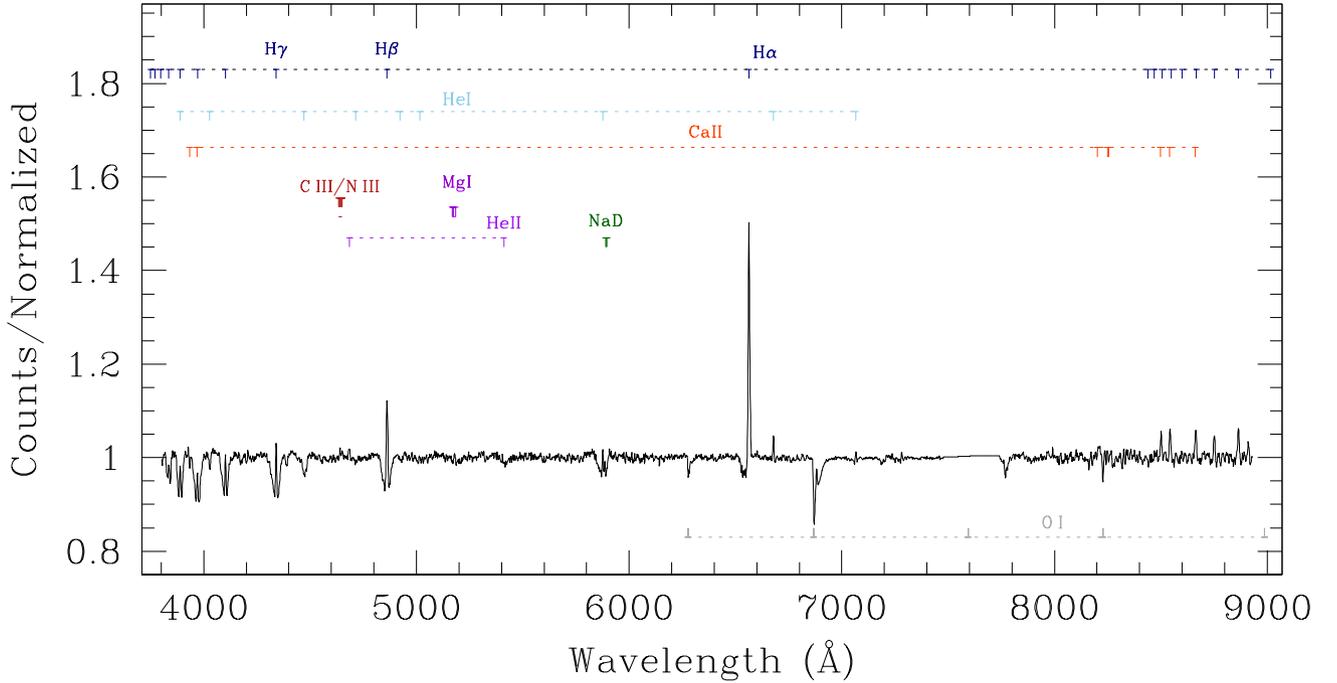}}
\end{picture}}
\caption{
 The combined and averaged low-resolution spectrum of \bg\  obtained at different epochs with different instruments. 
 Major lines identified in the spectrum are marked on top, while atmospheric oxygen lines are marked at the bottom.  \ion{Mg}{i} triplet is
 also marked, although we are not sure if it is real.}
\label{fig:lowres}
\end{figure*} 

\begin{figure*}
\setlength{\unitlength}{1mm}
\resizebox{15.cm}{!}{
\begin{picture}(130,75)(0,0)
\put(-15,0) {\includegraphics[width=8cm, bb=0 140 570 700, angle=0, clip]{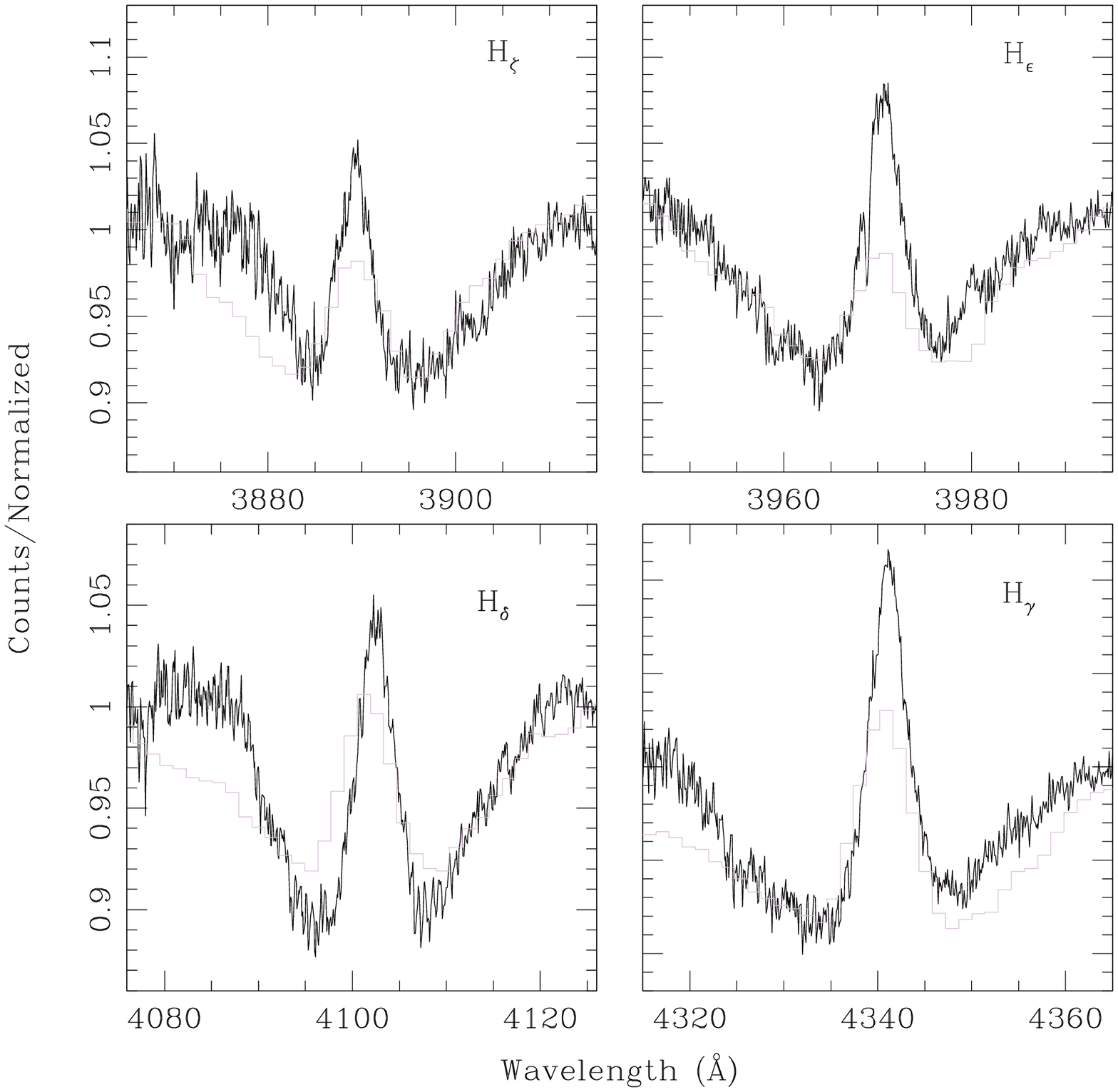}}
\put(60,0) {\includegraphics[width=8cm, bb=10 140 580 700, angle=0, clip]{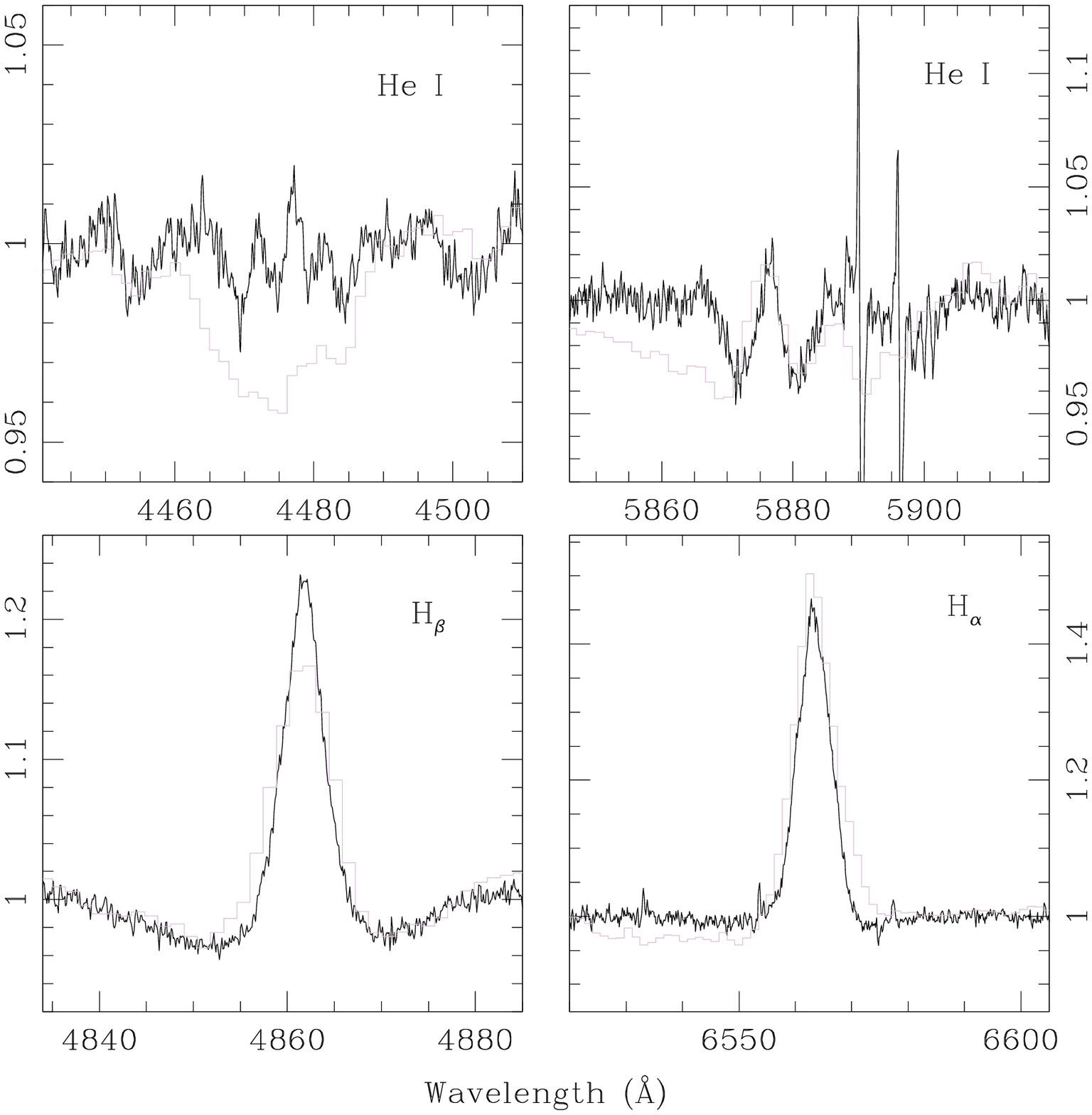}}
\end{picture}}
\caption{
 The profiles of important lines in the spectrum of \bg\ from the echelle spectra. The lines are marked in upper corners of each panels. The
 black line is the median of all high-resolution spectra,  whether the low resolution spectrum from Figure\,\ref{fig:lowres} are presented 
 in the background.}
\label{fig:lines}
\end{figure*}

The data from the Catalina Real-Time Transient Survey (CRTS) \citep{2009ApJ...696..870D} and the All-Sky Automated Survey for Supernovae 
(ASAS SN) \citep{2014ApJ...788...48S,2017PASP..129j4502K}  were used to produce the light curve  (Figure\,\ref{fig:lccurve}) of the object.
The data were obtained in $V$- and $g$-bands. The difference between filters $g$ and $V$ is not significant and the average colour index 
was $g-V\approx0.0$.
Most of the time  \bg\ is bright ($\approx$12~mag, one of the brightest CVs in apparent magnitude) and nearly constant with  non-regular, small amplitude variability. The average magnitude in a time stretch HJD 2456524.01 to HJD 2457668.08 is $V=11.91$, with a deviation 
of $\approx$0.06 magnitudes.  The light curve shows one occasion 
of a large flux drop ($\sim2.5$\,mag), with a duration of 176\,days. 
The initial descent from the average $V\approx11.9$ to $\approx$13.6 
magnitude lasts about 40\,days at a rate of 0.04\,mag~day$^{-1}$. 
Following a stand-still at that magnitude that last one month,  the brightness briefly falls further (on HJD\,$2458373$), reaching
$V=14.3$\,mag. From there, the luminosity starts to recover with a slower rate of 0.026\,mag\,day$^{-1}$.
This low state episode,  detailed in the upper panel of Figure\,\ref{fig:lccurve}, is fairly common among NLs, 
classifying BG\,Tri as a VY\,Scl star.
There is also an episode of a sudden jump of brightness detected between HJD\,2457669.88 and HJD 2457797.72 before 
the anti-dwarf nova episode when the average brightness of the object reaches $V\approx11.7$.

\subsection{Spectroscopy}

\subsubsection{Low-resolution spectroscopy}

A large number of low-resolution spectra were obtained with a variety of instruments listed in Table\,\ref{tab:speclog}. 
At the 2.5-m Isaac Newton Telescope (INT), we used the Intermediate Dispersion Spectrograph (IDS) with the R632V grating, 
the $2048\times 4100$ pixels EEV10a CCD detector, and a 1\farcs1 slit width. With this setup, we sampled the wavelength region 4400 - 6700~\AA\
at a  full width at half maximum FWHM of $\sim$2.5~\AA\ resolution.  
More spectra were obtained using the blue arm of the Intermediate dispersion Spectrograph and Imaging System (ISIS) mounted on the 4.2\,m
William Herschel Telescope, at the Roque de los Muchachos Observatory on La Palma. Spectra of Cu-Ne-Ar comparison lamps were obtained every
$\sim$30--40\,min in order to secure an accurate wavelength calibration. 

Additional data was obtained with the Andaluc\'ia Faint Object Spectrograph and Camera (ALFOSC), at 2.56m Nordic Optical Telescope (NOT) on La Palma. The $\#$8 CCD which has $2048\times2048$ pixel EEV chip was used for this observations. A spectral resolution of  FHWM $\sim 3.7$~\AA\  was achieved by using the grism $\#$7 (plus the second-order blocking filter WG345) and a 1" slit width. The useful wavelength interval this configuration provides is $\lambda$\,3800-6800~\AA.  The spectroscopy at the 2.2m Calar Alto  (CA) telescope was performed with CAFOS. A 1\farcs2 slit width and the G-100 grism granted access to the $\lambda$\,4200-8300~\AA\ range with a resolution of FWHM$\sim 4.5$~\AA\  on the standard SITe CCD ($2048\times2048$ pixels). 
After the effects of bias and flat field structure were removed from the raw images, the sky background was subtracted. The one-dimensional
target spectra were then obtained using the optimal extraction algorithm of \citet{Horne86}. For wavelength calibration, a low-order
polynomial was fitted to the arc data, the rms was always smaller than one tenth of the dispersion in all cases. The pixel-wavelength
correspondence for each target spectrum was obtained by interpolating between the two nearest arc spectra. The preliminary reduction steps  for all low-resolution spectra were performed with the standard packages for long-slit spectra within {\sc iraf}\footnote{IRAF is
distributed by the National Optical Astronomy Observatories, which
are operated by the Association of Universities for Research in Astronomy, Inc., under cooperative agreement with the National Science
Foundation.}, while wavelength calibration and most of the subsequent analyses made use of Tom Marsh’s {\sc molly}\footnote{{\sc molly} is
available at Tom Marsh’ web page: http://deneb.astro.warwick.ac.uk/phsaap/software/} package.

No flux calibration is available for the low-resolution spectra, hence we present in Figure\,\ref{fig:lowres} normalized spectrum
obtained by combining data at different epochs with different telescopes/instruments. The low-resolution  spectra failed to reveal 
significant radial velocity (RV)  variation. 

The overall spectral behaviour of 
\bg\ does not change significantly from epoch to epoch. The spectra indicate a steep blue continuum with Balmer lines showing emission features embedded in broader absorption lines. 
The higher members of the Balmer series appear to have  more intense absorption, while towards the lowest numbers the emission component
dominates. Helium lines are also present in the spectrum. The neutral helium lines 
have complex profiles, especially at \ion{He}{I} $\lambda4471$\,\AA. Also visible are \ion{He}{ii} and \ion{Ca}{ii} lines. 
The spectra are typical of NL variables  with an optically thick disc and low inclination.

\subsubsection{High-resolution spectroscopy}

The high-resolution observations were obtained with the echelle REOSC
spectrograph \citep{Levine1995}, attached to the 2.1m Telescope of the Observatorio Astron\'omico Nacional 
at San Pedro M\'artir, during several nights of 2017 and  2018. The 
CCD $2048\,\times\,2048$ detector was used to obtain a spectral resolution of R${\sim}18\,000$. All  observations were carried out with the
300~{l}~mm$^{-1}$ cross--dispersor, which has a blaze angle at around 5500~\AA. The spectral coverage was about $3600$--$7300 $ \AA. The
exposure time for each spectrum was 1200~s.  Th-Ar lamp was used for wavelength calibration. The spectra were reduced using the {\it
echelle} package in {\sc iraf}.   Standard procedures, including bias subtraction, cosmic-ray removal, and wavelength  calibration, were
carried out.
No flat-field correction and flux calibration was attempted; instead, the normalised spectra were used for measurements and visualisation.  The log of all 
spectroscopic observations is shown in Table~\ref{tab:speclog}.

\begin{figure}
\includegraphics[angle=0,trim=0 0 0 0,width=9cm]{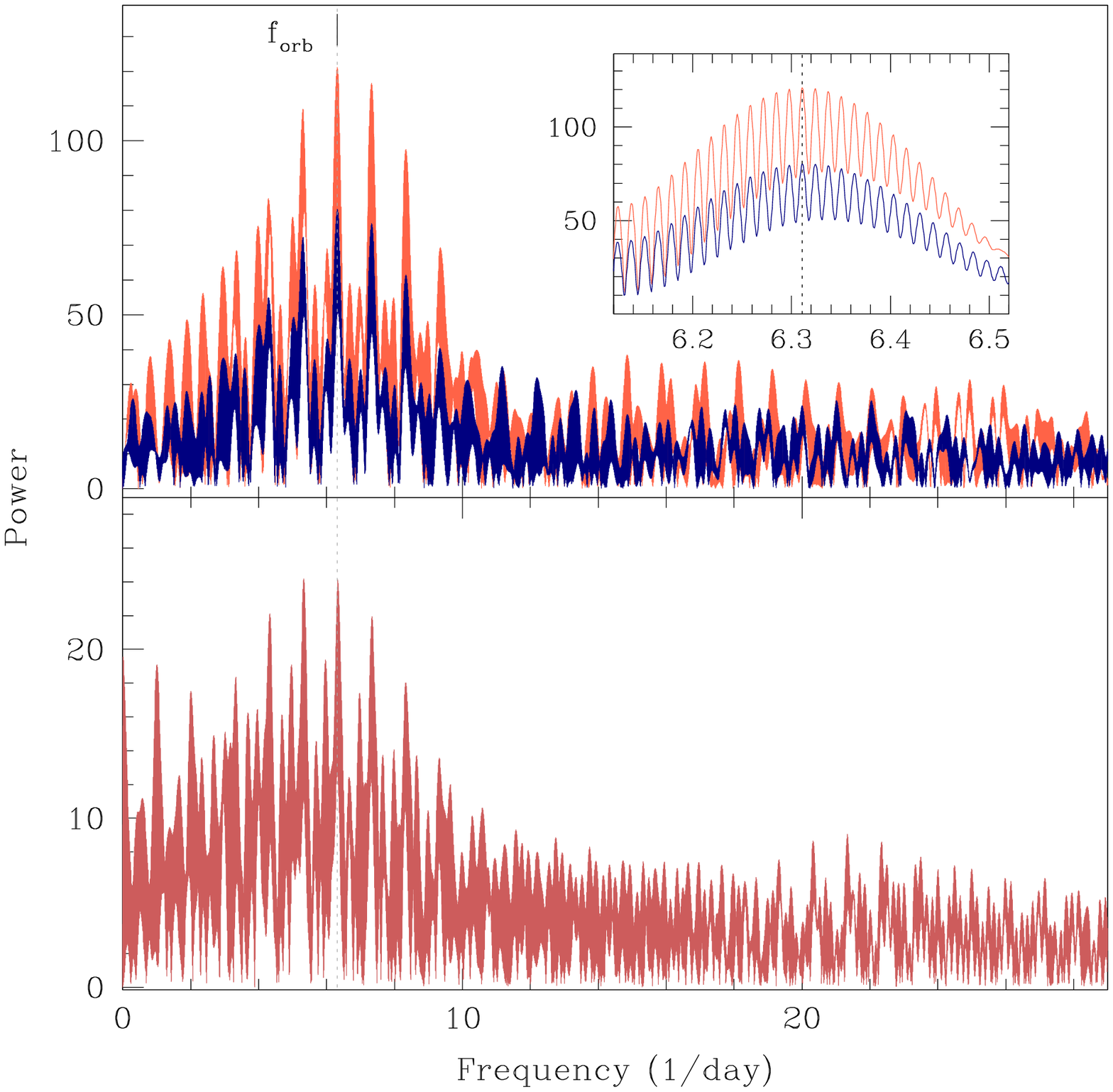}
\caption{
 The power spectra  of  \bg\ calculated for the  radial velocity curves of different lines and components.  In the upper panel the red and
 blue lines are the power of the HVC and the LVC after the  decomposition of  H$\alpha$ line. In the inset a zoom on the small range of
 frequencies around the derived period are presented. The bottom panel shows the power of H$\beta$  measured with a single Gaussian. }
\label{fig:power}
\end{figure}

\begin{figure*}
\setlength{\unitlength}{1mm}
\resizebox{15.cm}{!}{
\begin{picture}(130,60)(0,0)
\put(-10,0) {\includegraphics[width=15cm, bb=50 500 580 720, clip=]{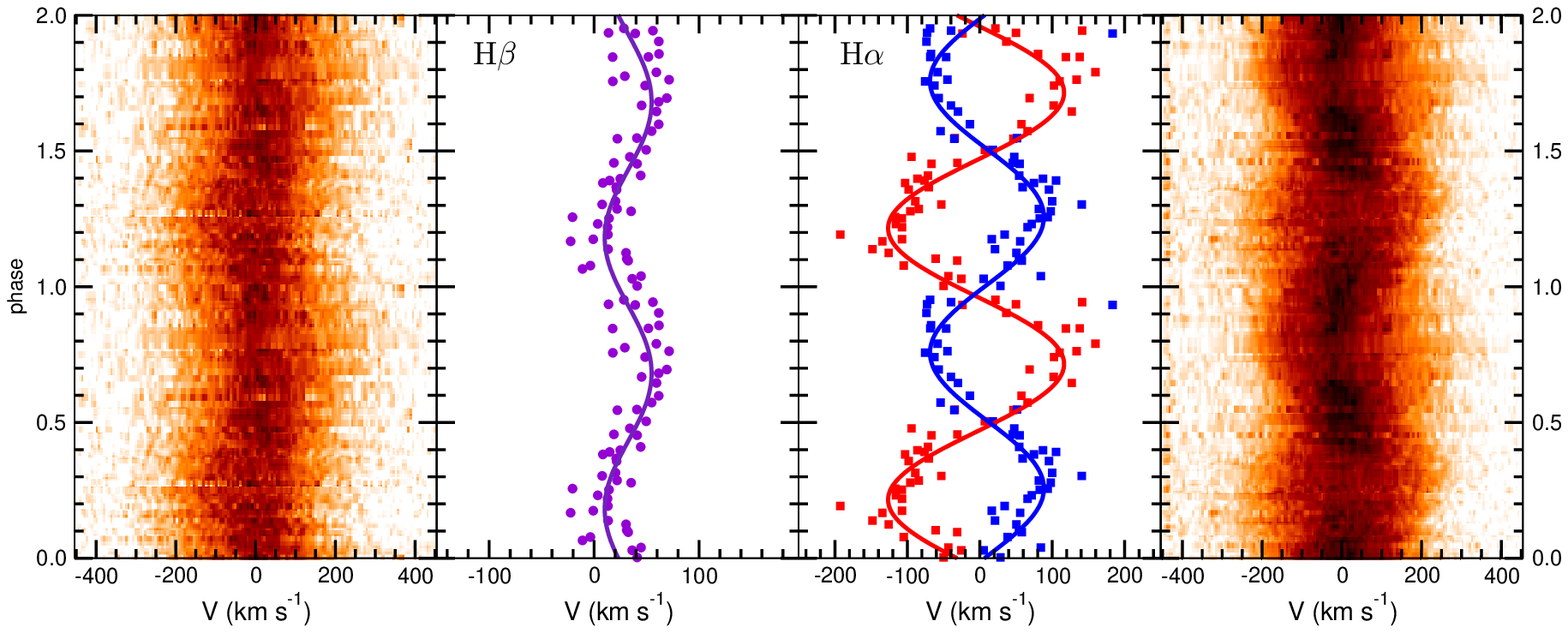}}
\end{picture}}
\caption{
 RV curves of H$\beta$ and components of H$\alpha$ folded with the orbital period are in the left and right panels respectively. RV of
 H$_\beta$ line was first measured by fitting a single Gaussian, which enabled determination of the orbital period.  H$_\alpha$ was separated
 into two components; by fitting two Gaussians (see the text for the description). The RV measurements corresponding to the high-velocity
 component of H$\alpha$ are marked by red symbols and the fit to them  with 121\,km~s$^{-1}$ semi-amplitude as a red line. Respectively, 
 the low-velocity component is plotted in blue. The semi-amplitude of the best fit curve is 74\,km~s$^{-1}$.  }
\label{fig:rvcurves}
\end{figure*}

\begin{figure}
\includegraphics[width=8.75cm,bb=20 150 580 700, clip=]{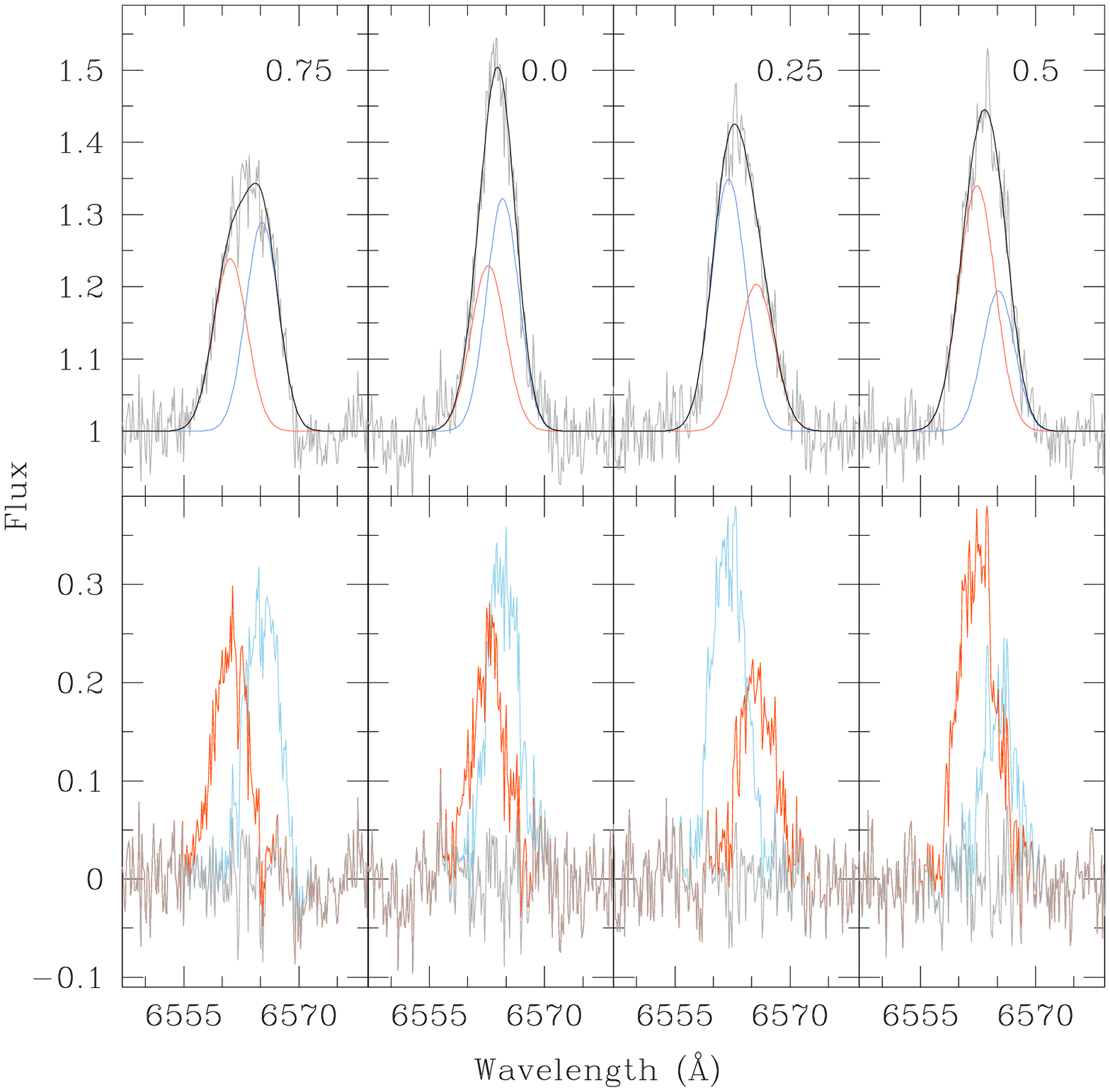}
\caption{
 The line profile phase-evolution of H$\alpha$. Four orbital phases (marked in the upper right corner of each panel) are displayed. In the
 upper panels the grey line are the observed spectra. Two Gaussians obtained as a result of de-blending are plotted by red and blue curves.
 Their sum is plotted by a black line as a fit to the observed profile. In the bottom panels by the red and blue lines the residuals of
 subtraction of the counterpart Gaussian are plotted, while the gray line is after subtraction of both Gaussians.}
\label{fig:profiles}
\end{figure}

\section{The phenomenology of \bg.}
\label{subsec:sed}

\citet{2008PZP.....8....4K} in a discovery note, points out that \bg\ is a variable star, probably a CV. Presented here, the long term light curve and spectra leave no doubt that the first assessment was correct. However, CVs comprised of a WD and a late red or brown dwarf secondary star come in different flavours depending on their orbital periods (or separation), mass accretion rate,  and strength of the magnetic field of 
the WD. Absence of nova or dwarf nova outbursts in a period of time over 5000\,days indicates that this is an NL variable. Moreover, an
anti-dwarf nova occurrence registered in the light curve is another characteristic of bright NLs, which occasionally undergo a fall in
brightness by more than one magnitude \citep{1995CAS....28.....W}. 

Low-resolution spectra confirm the CV identification of \bg, which exhibits a standard set of hydrogen and helium lines, with higher numbers
of the Balmer series showing wide adsorptions with embedded, relatively narrow emissions. In Figure\,\ref{fig:lowres}, an averaged and
combined spectrum obtained at different epochs is plotted. All significant spectral features are marked. 
A combination of emission and absorption features of Balmer lines  usually occurs  either in dwarf novae near period minimum  where the WD becomes dominant, or, in NLs with optically thick discs. 

Using the {\it Gaia} distance d= 334(8)\,pc to \bg,   we calculate the absolute magnitude of the object M$_V=4.26$. 
Such high luminosity enlists \bg\ among the brightest CVs, making it only the second after the infamous RW\,Sex \citep{2020MNRAS.496.2542H}.
The derived absolute magnitude is consistent with the  low inclination angle \citep{1986MNRAS.222...11W,1987MNRAS.227...23W}.  We will detail spectral energy distribution (SED) and luminosity of the accretion disc further in the paper. Nevertheless, based  only on the derived absolute magnitude, we can state that this object is not a low accretion rate WZ\,Sge type,  near the period minimum.

In NLs, the bulk of the accretion disc is optically thick (hot and dense), producing absorption lines.  The emission lines are formed either in the separate parts of the disc or by the gas in its vicinity \citep{warner_1976,1992A&A...256..433B}. Figure\,\ref{fig:lines} displays a set 
of \ion{H}{i} and \ion{He}{i} line profiles from averaged, and normalised high-resolution echelle spectra (with low-resolution profiles in 
the background), illustrating the composition of lines. In the averaged spectra, the emission profile of the Balmer lines  is   rather symmetric.
The absorption  is much wider,  visually slightly blue-shifted in regard to the emission peak. Measuring this shift in spectra not corrected for the instrumental sensitivity is complicated since the continuum is not well defined.

H$\alpha$ looks like it has a P\,Cyg profile with a blue-shifted absorption  from a material rapidly expanding in the direction of the
observer. However, it is just a contrast effect, particularly in the case of the low-resolution spectra. The emission feature is the 
strongest among the Balmer lines, hence the symmetry of the underlying absorption line is visually distorted. However, the composition of the 
line is similar to the rest of the Balmer lines; an emission peak is embedded in a wider absorption line.  
Helium lines show a more complicated structure comparison with Balmer lines, discussed below.

In general, \bg's spectrum looks like a replica of RW\,Sex \citep{2017MNRAS.470.1960H} or IPHASX\,J210204.7+471015 \citep{2018ApJ...857...80G}. Hence, our approach to interpreting  \bg\  builds on the assumption that it is a low inclination NL system.

\begin{figure*}
\setlength{\unitlength}{1mm}
\resizebox{15.cm}{!}{
\begin{picture}(130,100)(0,0)
\put(-10,0) {\includegraphics[width=13.5cm, bb=0 460 390 760, clip=]{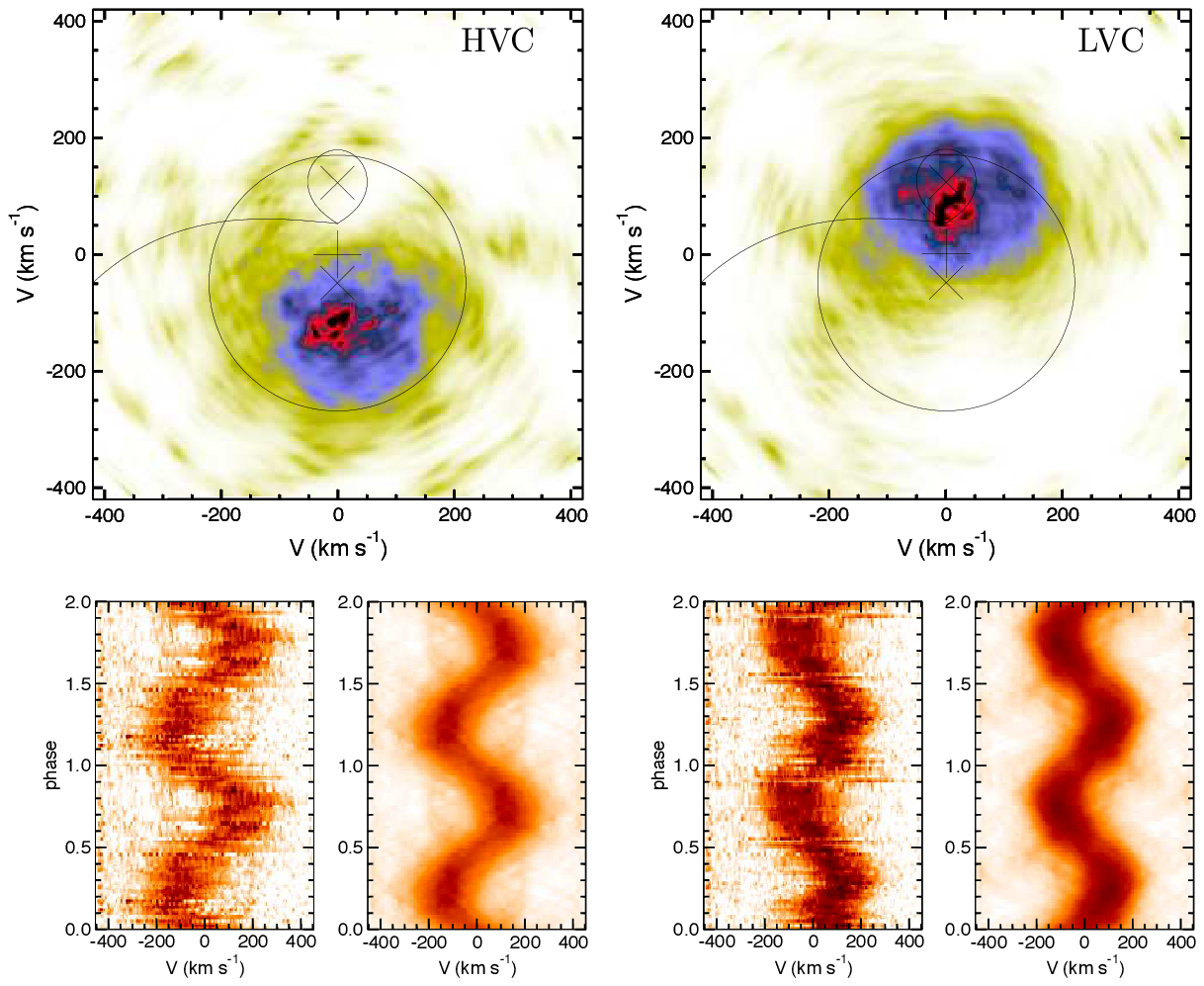}}
\end{picture}}
\caption{The trailed observed and reconstructed  spectra of H$\alpha$ components along with the Doppler maps are presented here. 
In the left side  the separated HVC trailed spectrum and its reconstruction are plotted at the bottom, while the corresponding velocity 
map of the HVC is displayed above. Respectively, on the upper right side of the figure is the Doppler tomogram of the LVC with the 
separated observed and reconstructed trailed spectra at the bottom.  
The Keplerian velocity of the disc in the Doppler maps is located at  $ \upsilon_{\mathrm {disc}}\sin (i) \ge 220$\,km\,s$^{-1}$ and 
the circle shows the disc external radius. The system parameters discussed in the text.
}
\label{fig:tomo}
\end{figure*}

\section{Radial velocities and period determination }
\label{sec:period}

Low-resolution spectra did not provide information on the orbital motion of emission lines.  This supports our assumption of a low orbital
inclination. Hence, we obtained higher resolution echelle spectra to reveal the orbital period of the object.   
Fitting H$\alpha$  emission with a single Gaussian produced a large scatter but no definite periodic pattern. 
However, measuring H$\beta$ with a single Gaussian produced a time series which allowed us to determine a possible period and its closest
one-day alias. The power spectrum calculated using a discrete Fourier transformation provided by  Period\,04
\citep{2014ascl.soft07009L,2005CoAst.146...53L} indicated that the orbital frequency was either $\sim$5.3~cycles-per-day (c/d) or
$\sim$6.3~c/d (see the bottom panel of Figure\,\ref{fig:power}). It was clear that the complications related to the period determination 
were due to lines that are composed of two or more components, as evident from the study of similar objects
\citep{2017MNRAS.470.1960H,2018ApJ...857...80G}.  

Now that we have an estimate of the orbital period, we  
further improve the obtained result by measuring individual components of the lines. 
In the high-resolution spectra, the H$\alpha$ line is the strongest feature. It is also the least affected by the absorption accompanying all H
emission lines. In general, it looks like a one peak line, but with variable wings. We applied line de-blending {\sl splot}-procedure in 
{\sc iraf} to separate the line into  components. 
We fit the profile with two Gaussians with unrestricted parameters.
The details of the  method are provided in \citet{2018ApJ...869...22T}.

We seek a periodic sinusoidal pattern in the RV/time-space by eye, and assign measurements of RVs to one or more
component(s), in all cases when they are clearly distinct. 
Sometimes just one component stands out.  
We determine a more accurate orbital period by subjecting the first emerging periodic pattern to the Fourier analysis. Period\,04 was
used to determine the best choice of the orbital period. We refine RVs by fitting a $sin$-curve to the measurements of the better-defined
component and perform another round of de-blending, by fixing the central wavelength of this component. As a result, we improve the
determination of the second component, which can be used to measure the orbital period. The difference in values of the orbital period
from two components is rather small.   The higher velocity amplitude component (HVC) power spectrum peaks at 6.31114\,c/d, while the lower
velocity amplitude  component (LVC) has a frequency of 6.33794\,c/d. The spread of measurements remains high. The residuals (RMS) of the fit of
HVC are 36\,km~s$^{-1}$ with the amplitude of variation 121\,km~s$^{-1}$. The LVC has an amplitude of 78\,km~s$^{-1}$ and
RMS of 25\,km~s$^{-1}$.  They reconcile better at the period determined from the HVC, which we adopted as the orbital period of 
P$_{\mathrm {orb}}=3.8028(24)$\,h = 0.15845(10)\,d.
The power spectra obtained from RVs of H$\beta$ and two components of H$\alpha$ are presented in Figure\,\ref{fig:power}. Obviously, the power
of H$\beta$ as a whole is much smaller than the power of separate components of H$\alpha$, but they all are consistent with one another.

The RV curves of H$\beta$  and H$\alpha$ components along the measurements folded with the determined orbital period are presented in the two
middle panels of Figure\,\ref{fig:rvcurves}, respectively. We assigned phase zero at the negative to positive crossing of the LVC.  Plots are
flanked by trailed spectra of corresponding lines on both sides.
In the upper panel of Figure\,\ref{fig:profiles} Gaussian profiles of individual components, their sum, and the observed line profiles in four
different orbital phases are presented to illustrate the result of the separation of the line  into two components. Residual spectra of individual components,  as well as residuals after subtraction of both components, are displayed in the bottom panel.
H$\beta$ is probably also multi-component, but dismantling  it into components is rather difficult. There is a stronger absorption
undermining the wings of the emission components. One component (the HVC) dominates, as evident from Figure\,\ref{fig:rvcurves}. The amplitude of fitted $sin$-curve to single-Gaussian measurements of the line is only 24\,km\,s$^{-1}$, decreasing to 22\,km~s$^{-1}$ if the adopted 0.15845\,d
period is used to fit the data. The RMS is 16\,km~s$^{-1}$.

\begin{figure}
\setlength{\unitlength}{1mm}
\resizebox{15.cm}{!}{
\begin{picture}(130,105) (0,0)
\put(0,0) {\includegraphics[width=13cm, bb=0 500 390 820, clip=]{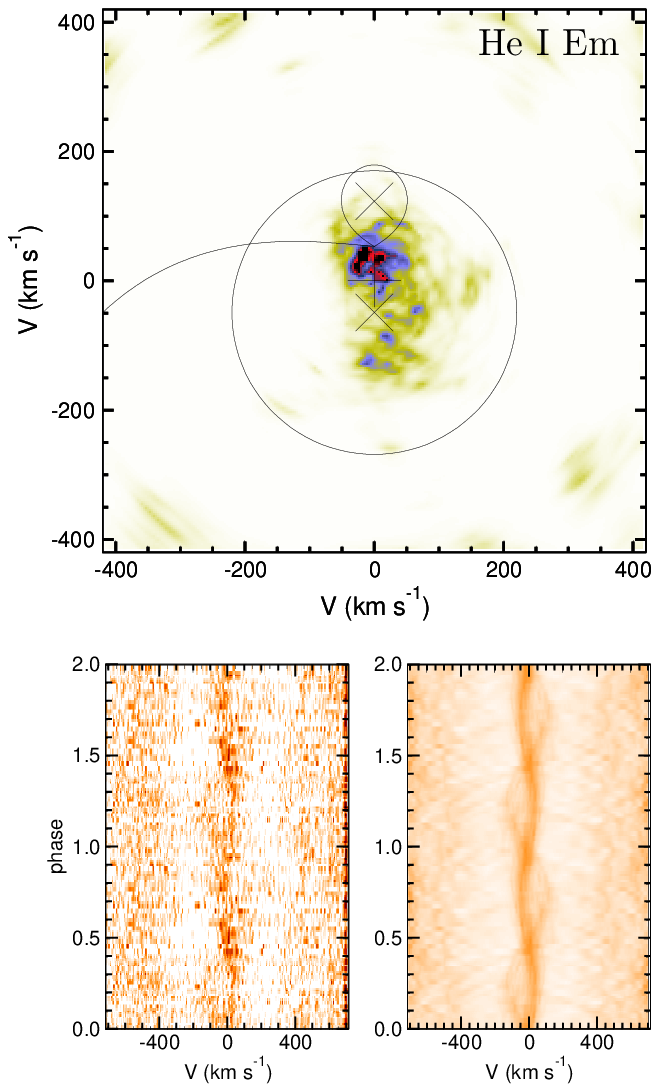}}
\end{picture}}
\caption{The Doppler map of He I $\lambda5875.6$\,\AA\ emission. The trailed spectrum and reconstructed trailed spectrum of
the emission are presented in the bottom and the velocity map on top panel. 
The Keplerian velocity of the disc in the Doppler maps is located at $ \upsilon_{\mathrm {disc}}\sin (i) \ge 220$\,km\,s$^{-1}$ and the circle
shows the disc external radius. The system parameters discussed in the text.}
\label{fig:tomohei}
\end{figure}

Armed with a set of two components comprising  H$\alpha$,   we may try to figure out their origin. 
We compare our spectroscopic observation with a small sample of similar objects published recently
\citep{2017MNRAS.464..104H,2017MNRAS.470.1960H}, which show  similar two-component emission lines. 
RW\,Sex and 1RXS\,J064434.5+334451 \citep{2017MNRAS.470.1960H} and  RW\, Tri \citep{Subebekova20} were all observed with the same instrumental
settings as \bg, so the measurements are uniform, and the comparison is straightforward. All three objects show two distinct components
varying with the respective orbital periods in almost counter-phase, relative to one another. One component is usually wider.
Another, the narrow one, is firmly linked to the irradiated face of the secondary by two eclipsing objects in the sample, for which the zero
phases were known precisely.
The wide component is also regularly the higher velocity component in other NLs. In case of \bg, the difference in widths of components is
insignificant, but still, the HVC is slightly broader than LVC (average FWHM$=5.7(8)$\,\AA\ vs $5.3(4)$\,\AA,  respectively). 

Assuming that the secondary star emits the LVC, in analogy with the above-mentioned NLs \citep{2017MNRAS.470.1960H,Subebekova20}, we determine the orbital phase zero ($\upvarphi=0$)
of the system as the moment when the RV of LVC changes sign from negative to positive, i.e. when the secondary star is in the inferior
conjunction. In which case, the ephemeris  of \bg\ can be expressed as 
\begin{equation}
 \mathrm{HJD_{\upvarphi=0}} = 245 8053.45490(60) + 0\fd15845(10)\times\mathrm{E}.
 \end{equation}
All phases used in this paper were calculated with this ephemeris. 

\begin{figure*}
\setlength{\unitlength}{1mm}
\resizebox{15.cm}{!}{
\begin{picture}(130,100)(0,0)
\put(-10,0) {\includegraphics[width=15cm, bb=35 95 530 430, angle=0, clip]{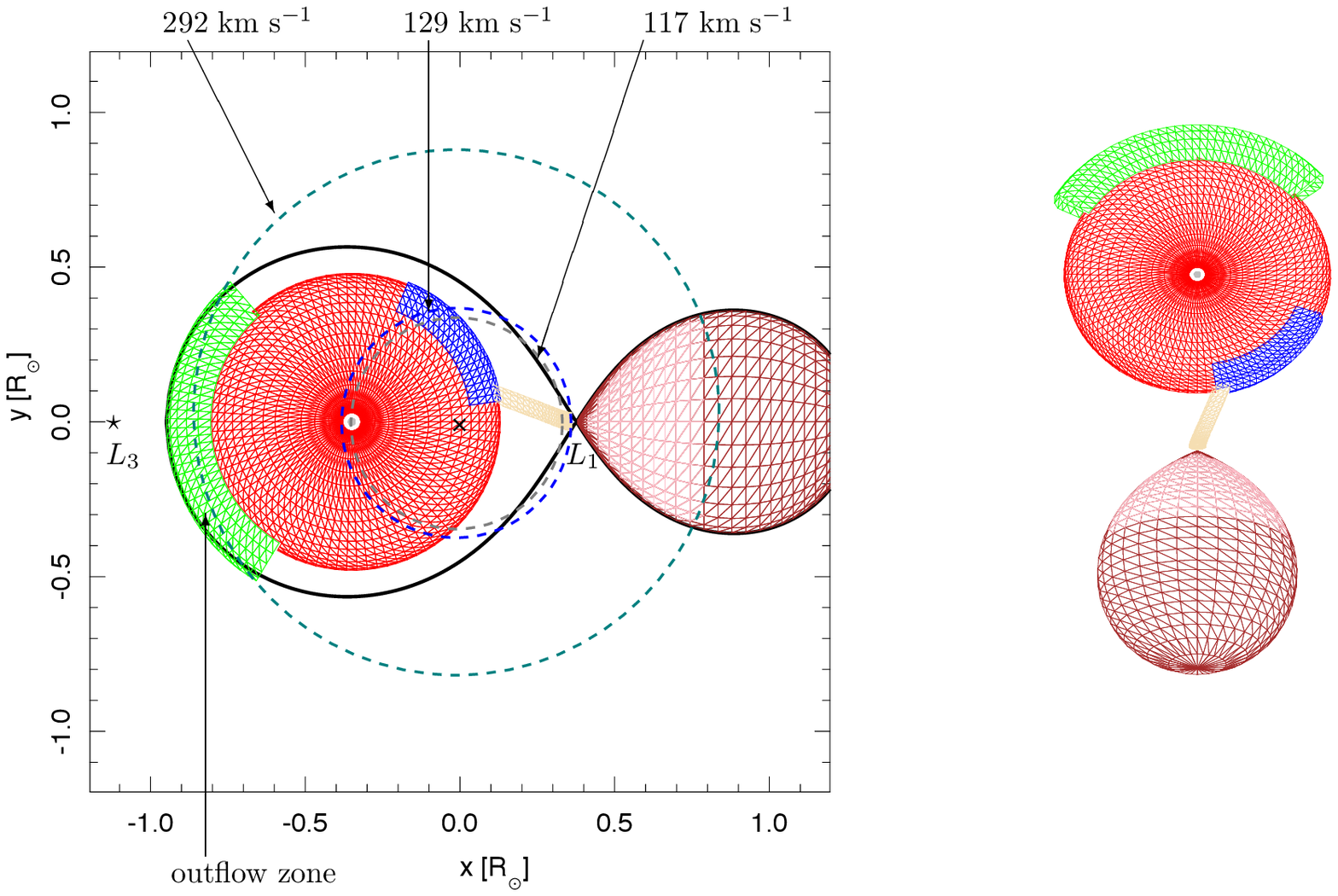}}
\end{picture}}
\caption{The geometric model of BG Tri.  The center of mass (x) and Lagrangian corresponding points are marked.  Velocities of the  L$_1$ point,  the disc outflow area (green), and the WD are denoted. LVC which is emitted from the irradiated hemisphere of the secondary star has a velocity exceeding that of the L$_1$ point and less than the orbital velocity of the star.}
\label{fig:Model}
\end{figure*}

\subsection{Doppler tomography}
\label{subsec:doptom}

A customary  way to illustrate the emission line behaviour is via trailed spectra and Doppler tomography \citep{1988MNRAS.235..269M}. It is
worth mentioning that the Doppler tomography works best if the emitting particles are in the orbital plane, and the inclination of the
system's orbital plane is high \citep{2005Ap&SS.296..403M}.  
In order to construct trailed spectra and Doppler maps, we used the phase zero as defined above, stemming from the assumption   
that the LVC is the component originating from the irradiated face of the secondary.
With this premise, the Doppler tomograms presented in Figures\,\ref{fig:tomo} and \ref{fig:tomohei} were constructed
\citep{1998astro.ph..6141S}. Echelle high-resolution spectra were used for this purpose. 
 The Doppler map of the H$\alpha$ line without separation into two components was calculated, but is not presented here because it is less informative. Tomograms of unaltered lines were presented by \cite{2017MNRAS.470.1960H,2018ApJ...857...80G}. Instead,  we split the  H$\alpha$ line into components, as demonstrated in  the bottom panel of Figure\,\ref{fig:profiles}, and made a tomogram of each component separately.

In the bottom panels of Figure\,\ref{fig:tomo} trailed spectra (and their reconstructed counterparts) of
H$\alpha$'s  HVC and LVC are presented side by side. Together, they form 
the trailed spectra of H$\alpha$ shown at the right side of Figure\,\ref{fig:rvcurves}.  Meanwhile,  in the upper panels, the velocity maps
resulting from the Doppler tomography are shown.  Locations of the
stellar components are marked by "$\times$", the centre of the masses by "$+$". The Roche lobe of the secondary star and the
ballistic trajectory of mass transfer flow are over-plotted, as well as a ring corresponding to
the outer radius of the disc. To calculate them, we selected the mass of the WD, M$_{\mathrm{WD}}=0.8$\,\msun, and the mass
ratio,  $q=0.4$\  as statistically average values for a P$_{\mathrm {orb}}=3.8$\,h nova-like CV
\citep{2011ApJS..194...28K,2011A&A...536A..42Z}.
The orbital inclination angle was fixed at  $i=25$\grad\ (see Section\,\ref{systempar} for justification of these parameters). 

The LVC produces a spot converging with the position of the secondary in the velocity map. Apparently, this is a result of our phase
allocation. Usually, when the irradiated secondary star forms the emission, the line is narrower  (3.8\,\AA\ in the eclipsing RW\,Tri
\citep{Subebekova20} against 5.3\,\AA\ in \bg), and the compact spot is concentrated at the hemisphere facing the WD. The reason for such a
diffuse spot is not clear, but a low orbital inclination of the system probably contributes to it.  
Meanwhile, the HVC produces another diffuse  and elongated concentration (upper left panel), at a region which is clearly evading
identification with either the stellar component of the binary, the accretion disc, or the mass transfer stream; including impact area of the
stream with the disc. Detection of this H$\alpha$ component and its corresponding location in the velocity maps has become common for
RW\,Sex-type NLs. Among a few possible explanations cited by \citet{2017MNRAS.470.1960H} and \citet{Subebekova20}, we prefer the 
model in which the outflow from the disc takes places in the orbital plane, instead of the wind perpendicular to the plane direction. Three-dimensional numerical simulations of the gas dynamics show that such outflows are viable through the vicinity of the Lagrange L$_3$ point \citep{2007ARep...51..836S}. \bg\ provides a good argument in favour of this hypothesis: firstly, the object shows meagre emission from \ion{He}{ii} line, even though we observe it almost face-on, hence it is difficult to argue that a disc wind 
is significant enough to produce an intense enough emission spectrum \citep[e.g.][]{10.1093/mnras/stv867}  to overcome the bright accretion
disc. Generally, P\,Cygni-like profiles for NLs are observed in UV, and quite successfully modelled for recombination emission
\citep{2002ApJ...579..725L}, where the lines of low inclination systems  exhibit the characteristic imprint. However, reproducing it in the
optical range is a challenging task; it produces some double-peaked emission line contribution only in a high inclination ($i>60^\circ$) systems.
\citet{2016PhDT.......348M} attempted to convert lines into the single peaked, requiring an extension of the line forming region to $\sim$150
R$_{\mathrm {WD}}$, which again works only for high inclinations. As a result, \citet{2016PhDT.......348M} succeeded to produce a successful
model for RW\,Tri by comparing it to a low-resolution spectrum of the object. But \citet{Subebekova20} demonstrated that the H$\alpha$ in RW\,Tri
is complex and at least partially formed at the heated surface of the secondary star.   
Secondly,  H$\alpha$ emission has been disentangled already in four objects, including \bg,  all of which have vastly different inclination
angles. After correction for the inclination angle $i=25$\grad (see Section\,\ref{systempar}),  the measured $\upsilon_{\mathrm{HVC}}/
\sin(i)=121$\,km~s$^{-1}$ of the HVC  converts into  286~km~s$^{-1}$. According to \citet{2017MNRAS.470.1960H} and \citet{Subebekova20} the
HVC  (also called "the wide component" in these publications) in all studied objects is $\approx300$\,km~s$^{-1}$, after the inclination angle
is accounted for. 
The fact that the HVC in all morphologically similar NL variables obtains the same value  after correction for a variety of inclination
angles, is a good evidence that the source of HVC is in the orbital plane. 
One would expect that the wind will constitute itself  differently, depending on the viewing aspect.  Notwithstanding, the HVC always appears on the
Doppler maps at the same spot, adjacent to the outer edge of the accretion disc on the opposite side from the secondary/hot-spot,
regardless of the system inclination. 

Figure\,\ref{fig:lines} demonstrates that \ion{He}{i} acts somewhat differently with respect to the Balmer lines. It is also a common
characteristic of RW\,Sex-type NLs. 
Helium lines are  fainter than H$\alpha$ or H$\beta$ and are comprised of equally intense emission and absorption. Nevertheless, we explored
their phase evolution using illustrative trailed spectra and their respective tomograms. We concentrated on the \ion{He}{i}
$\lambda\,5875.6$\,\AA\ line, which has the best S/N among helium lines. A trailed and reconstructed spectra of the emission component of this  line,  and the corresponding tomogram
is presented in 
Figure\,\ref{fig:tomohei}. Meanwhile, the same line is treated as an absorption feature  on the right side.
The emission component of the line basically repeats the pattern observed in H$\alpha$. The observed trailed spectrum is faint, but the
program picks up the signal correctly, reflecting it in the reconstructed spectrum. The low-velocity component definitely dominates, and hence
a compact spot materialises  close to the L$_1$ point, contrary to the diffuse appearance in the  H$\alpha$ case. The HVC is hardly traceable
in the observed trailed spectra, but apparently some emission comes from here too. Hence, a coma-like extension in the velocity map. 
%

\begin{figure}
\setlength{\unitlength}{1mm}
\resizebox{15.cm}{!}{
\begin{picture}(130,55)(0,0)
\put(0,0) {\includegraphics[width=7.25cm, bb=15 200 585 630, clip=]{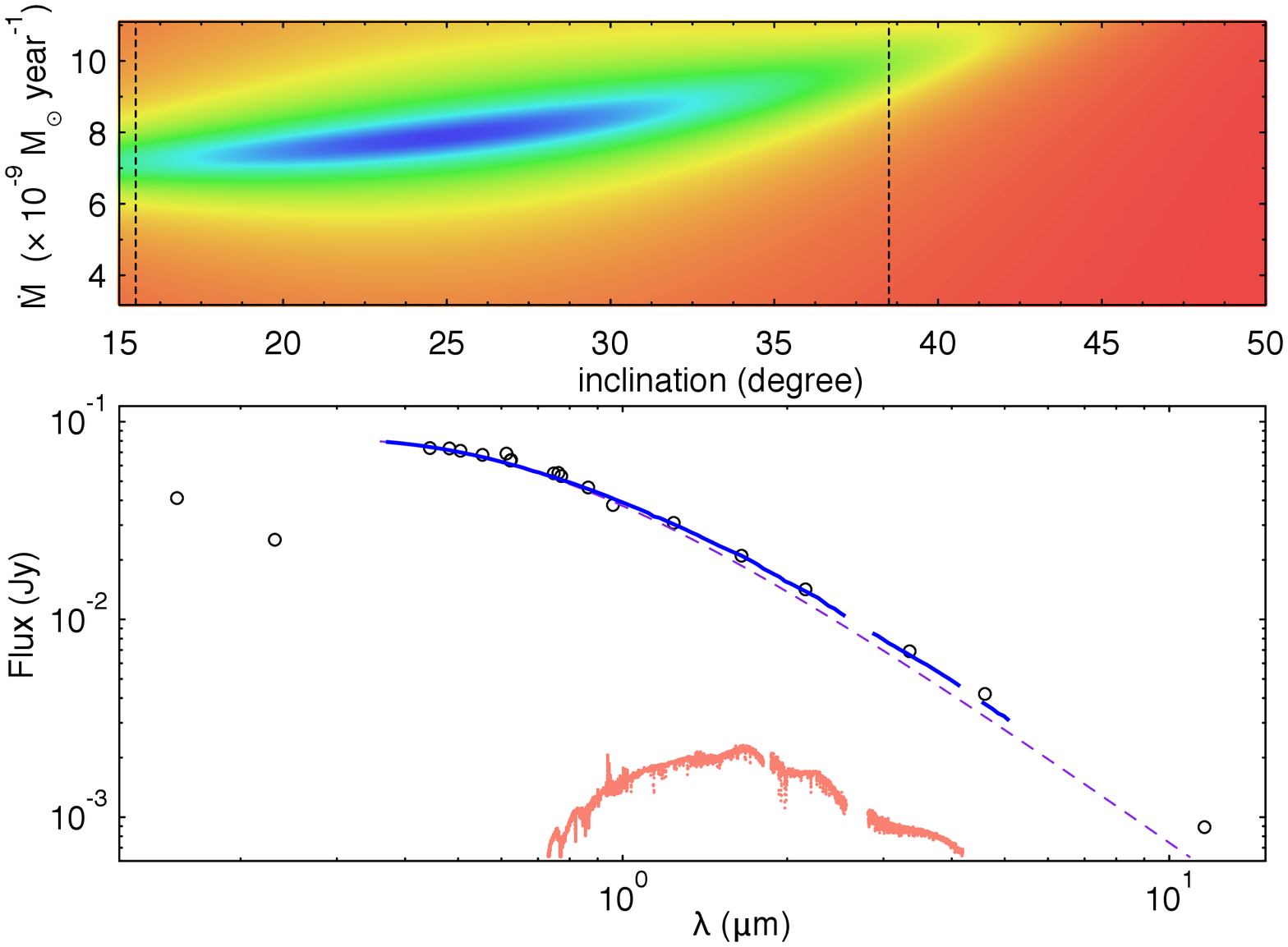}}
\end{picture}}
\caption{
 The plot of spectral energy distribution of \bg\ in the bottom panel. Circles represent the fluxes from 
 Table\,\ref{tab:SEDlog}. They were corrected for the interstellar extinction E($B-V$)=0.03.
 The dashed line through the observed data in the optical-IR range  represents the spectrum of accretion disc.   
 The red points represent a spectrum of a M3V star scaled to  a 337\,pc distance.
 The solid line is a combination of fluxes from the accretion disc and the secondary. The outermost left two points correspond to UV data. 
 In the upper panel the best fit parameters (mass accretion rate vs. inclination angle) of the accretion disc model to the data are presented
 in the form of intensity map. The extreme boundaries of $i$ corresponding to the secondary mass center and L$_1$ point are marked by
 vertical lines. Preferred values of \mdot and $i$ (blue strip) are also function of increasing surface and decreasing temperature along
 latitude. }
\label{fig:sed}
\end{figure}

\section{SYSTEM PARAMETERS}
\label{systempar}

The average masses of WDs in CVs   above the period gap is M$_{\mathrm{wd}}$(P$_{\mathrm{orb}} > 3$ h) $= 0.86(20)$ \msun\
\citep{2011A&A...536A..42Z}.
For simplicity, we adopted  M$_{\mathrm{WD}}=0.8$\,\msun. Taking into account the semi-empiric relation of secondary type vs orbital period, we
expect a secondary with a mass of M$_2\approx 0.3$\,M$_{\sun}$, and spectral type of M3V-M4V \citep{2006MNRAS.373..484K}. The expected mass ratio
and separations are $q \equiv {\mathrm{M}}_2/{\mathrm{M}}_{\mathrm{WD}}\approx0.4$ and $a=1.23$\,R$_{\sun}$, respectively. The  accretion disc
truncation radius\footnote{equation 2.61 from \citet{1995CAS....28.....W}} is $r_{\mathrm{disc}}^{\mathrm{out}} = 0.52$\,R$_{\sun}$.
Corresponding primary, secondary, and L$_1$ orbital velocities are 117\,km s$^{-1}$,  292\,km\,s$^{-1}$, and  129\,km\,s$^{-1}$, respectively (see
Figure~\ref{fig:Model}).

 There are numerous flux measurements of \bg\ available  in the public domain. We fetched all available data from
 VizieR\footnote{http://vizier.unistra.fr/vizier/sed/} \citep{2000A&AS..143...23O}.
 We revised the available data, since some measurements are erroneous  (e.g. the SDSS data are not correct because the object is too bright),
 and  compiled those which passed the scrutiny in Table\,\ref{tab:SEDlog}. 
Selected data are plotted using  circles in the bottom panel of Figure~\ref{fig:sed}, after the interstellar reddening correction
of E($B-V$)=0.03 \citep{2015ApJ...810...25G}.
Using the  {\it Gaia} distance of $d= 334(8)$\,pc, we can calculate the luminosity of different components.
In particular, even a hot 50\,kK white dwarf will have a negligible contribution to the optical flux. A secondary M3-4\,V Roche lobe filling
star has some insignificant influence in the IR. The comparison of the observed fluxes with those expected from different components confirms that
the flux in the entire wavelength range  is formed mostly by the accretion disc.  
The dashed lines in the bottom panel of Figure~\ref{fig:sed} indicate a simple accretion disc spectrum as a composition of multiple
black-bodies; from concentric rings with a corresponding distribution of temperatures throughout the stationary, optically thick disc
\citep{1989A&A...211..131L}. Since the observed flux would depend on the aspect of the disc, we fitted the observed SED as a function of the
mass transfer rate \mdot, and  the system inclination.

The UV points were excluded from consideration because  it is well known that accretion disc models do not reproduce  the observed spectrum
shape of CVs in the wide range including the UV \citep{2007AJ....134.1923P}. 

For the fit's robustness, we added a  spectrum of  M3V-type stars from the empirical template library of the Sloan Digital Sky Survey stellar
spectra \citep{Kesseli:2017aa}, which was extended to the IR range using  spectral templates from \citet{Rayner:2009aa}.
The star spectrum was scaled to the object's distance.  For each band listed in Table\,\ref{tab:SEDlog} (except $FUV$ and $NUV$) the flux $m_{calc}$ was calculated as a sum of the disc model and the secondary.
The result of the best fit of function $\chi^2 = \Sigma ((m_{obs} - m_{calc})/\Delta m_{obs})^2$ is presented as a long dashed line in the
lower panel of  Figure\,\ref{fig:sed}. 
In the upper panel of  Figure\,\ref{fig:sed}, the goodness-of-fit is presented as an intensity scale diagram. 
Following the assumption that the LVC is emitted from the heated hemisphere of the companion star, the expected LVC velocity is located in the
range of  $\upsilon_{\mathrm{LVC}} \in $  [130\,$\sin(i), 290\,\sin(i)$\,km~s$^{-1}$] (see Figure\,\ref{fig:Model}), which allow us to estimate
limits on the orbit inclination from the dynamical constrains.  The observed value of LVC is 78\,km~s$^{-1}$, which defines the limits marked by
the vertical dashed lines in  Figure\,\ref{fig:sed}, top. 
The line at the right side of the plot indicates the observed LVC, which is consistent with the L$_1$ point, and another corresponds to the center of
mass of the secondary (at the left side). In other words, the inclination angle and mass accretion rate would be highest if the LVC is
emitted just from the  L$_1$ point, and accordingly, lowest if the entire surface of the secondary is heated. Of course, neither
assumption is correct,
so we introduced some parameter which reflects the increasing cross-section of the secondary, with decreasing temperature. According to which, 
the best fit is achieved  at $i=25(5)$\,degree and \mdot=$8.0(1.0)\times10^{-9}$\,\msun\,year$^{-1}$.

Given all uncertainties of adopted assumptions, the fit is remarkably good. The values of mass accretion rate are within the range of
estimates for a number of other NLs \citep[e.g. Figure 2 of ][]{2019arXiv191001852H}.  The mass transfer \mdot\  dependends slightly on
the mass of the WD. For a 0.6\,\msun WD,  the mass transfer rate  increases to  10$^{-8}$\,\msun\,year$^{-1}$. On the down side,   \mdot\  is
$\sim4.0\times10^{-9}$\,\msun\,year$^{-1}$  for a massive $\geq1.0$\,\msun\ WD.

 The LVC velocity corrected for a system inclination of  $i=25$\,\grad\  is $\upsilon_{\mathrm{LVC}} =$ 185\,km~s$^{-1}$. 
As we showed earlier, the corrected HVC velocity is $\approx300$\,km~s$^{-1}$,   it corresponds to the orbital velocity at the edge of the disc on the opposing side of the system relative to the secondary.
The location of the HVC on Doppler maps of all similar NLs is related to the Lagrangean  L$_3$ point. According to some hydrodynamic models
\citep[]{2007ARep...51..836S, 2017MNRAS.467.2934L}
an outflow of matter from the disc  in to the orbital plane takes place from this area (marked green area in
Figure\,\ref{fig:Model}). The concentration of gas in that area might be responsible for HVC or wide component of emission lines in these
objects.

\begin{table}

\caption{VizieR photometric data of BG Tri}
\label{tab:SEDlog}
\begin{tabular}{cccccc}
  \hline \hline
$\lambda$  & Flux & 1$\sigma$ Flux & Band & Source & Ref. \\
($\mu$m)  & ($\times 10^{-3}$Jy)  & $\times 10^{-3}$   &    &  &   \\ \hline
0.153 & 32.1 &0.2  & $FUV$ & \textit{GALEX} & 1 \\
0.231 & 19.3 &0.1  & $NUV$ & \textit{GALEX} & 1\\
0.444 & 64.2 &	11.6	& :B & AAVSO  & 2 \\
0.477 & 79.0 & 	    & $g$  & PAN-STARRS & 3 \\
0.482 & 64.6 &	9.3 &	$g'$ & AAVSO & 2 \\
0.505 & 65.7 & 0.9 & $Gbp$ & \textit{GAIA2} & 4 \\
0.554 & 61.4  & 9.6 & $V$ &	AAVSO & 2 \\
0.613 & 63.2 & 	   & $r$ & PAN-STARRS & 3 \\
0.623 & 58.9 & 0.3 & $G$ & \textit{GAIA2} & 4 \\
0.625 & 58.7 &	7.7& $r'$ &	AAVSO & 2 \\
0.748 & 51.6 & 	   & $i$ & PAN-STARRS & 3 \\
0.763 & 51.5 &	7.6 &	$i'$ & AAVSO & 2 \\
0.772 & 49.6 & 0.7 & $Grp$  & \textit{GAIA2} & 4 \\
0.865 & 44.2 & 	   & $z$ & PAN-STARRS & 3 \\
0.960 & 36.5 & 0.8& $y$ & PAN-STARRS: & 3 \\
1.24 & 29.9 &0.6 & $J$ & 2MASS & 5\\
1.65 & 20.6 &0.4 & $H$ & 2MASS & 5\\
2.16 & 14.0 &0.2 & $Ks$ & 2MASS & 5 \\
3.35 & 6.85 &0.15 &$W1$ &  \textit{WISE} & 6\\
4.60 & 4.19 &0.08 & $W2$ &  \textit{WISE} & 6 \\
1.16 & 0.89 &0.09 & $W3$ &  \textit{WISE} & 6 \\
 \hline
\end{tabular}
\begin{tabular}{l}
1 - \citet{2017ApJS..230...24B} \\
2 - \citet{2015AAS...22533616H} \\
3 - \citet{2016arXiv161205560C} \\
4 - \citet{2018AA...616A...1G} \\
5 - \citet{2003yCat.2246....0C} \\
6 - \cite{2012yCat.2311....0C} \\
\end{tabular}
\end{table}

\section{Conclusions}
\label{sec:conclusions}

We studied \bg, one of the brightest CVs at $V=11.9$, which somehow escaped attention until now. We determined its orbital period to be
0.15845\,d or 3.8028\,h.  We show that it is a  NL system, identified by its characteristic blue spectrum containing a set of \ion{H}{i} and
\ion{He}{i} lines, comprised of wide absorption features containing strong emission peaks. The orbital period and the spectrum, combined with
the SED, indicate the presence of a bright accretion disc in a high density and temperature state, proper to NLs. The long term light curve of
\bg, obtained by ASAS\,SN and CRTS sky surveys, shows the absence of dwarf novae style outbursts, but reveals an instance of low luminosity
state often detected in NLs, also known as the VY\,Scl phenomenon.

 Absorption lines originate from the optically thick accretion disc, while emission forms elsewhere.  We demonstrate that the Balmer emission
lines are complex and we are able to separate H$\alpha$ into two components.       
We identify the LVC with the heated surface of the secondary star facing the luminous disc. 
We associate the HVC  with the disc outflow region situated on the opposite from the
secondary and the hot-spot, side of the disc. A similar occurrence is also common for mentioned NLs, which we may call RW\,Sex type systems, all of which are concentrated in a 3-6\,h range of orbital periods \citep{Subebekova20}. 
An HVC velocity, corrected for the inclination angle, of $\sim$300\,km sec$^{-1}$, is very definite in all studied objects, regardless of their
orientation. That, in our opinion, argues against the disc wind origin of the HVC component. However, it is not evidence of the absence of wind,
just a rationalisation that emission lines in the optical range are not formed in the wind.

The SED of \bg\ confirms that stellar components contribution is negligible and that most of the flux from UV to  near-IR is emitted by the
disc. The energy balance favours  that we observe the system nearly face-on, and the deduced inclination angle validates the value fetched
from the dynamical considerations.

\section*{Acknowledgements}

 We are grateful to the anonymous referee for the valuable comments which helped to improve this paper. This work is based upon observations carried out at the OAN SPM, Baja California, M\'{e}xico. We thank the daytime and night support staff at the OAN-SPM for facilitating and helping obtain our observations. This research has made use of the VizieR catalogue access tool, CDS, Strasbourg, France (DOI: 10.26093/cds/vizier). The original description of the VizieR service was published in A\&AS 143, 23.
GT and SZ acknowledge PAPIIT-DGAPA-UNAM (grants IN108316, IN102120 and IN110619) and CONACyT grant 166376.  MSH acknowledges the Fellowship for National PhD from ANID, grant number 21170070. MSH and GT are thankful to SIMA project 687 of UNAM.   This research has been was funded in a part by the Science Committee of the Ministry of Education and Science of the Republic of Kazakhstan (Grant No. AP08856419).
BTG was supported by a Leverhulme Research Fellowship and the UK STFC grant ST/T000406/1. AA received support from Thailand Science Research and Innovation (TSRI) grant FRB640025 contract no. R2564B006.

\section*{Data Availability}
The data underlying this article will be shared on reasonable request to the corresponding author.


\bibliographystyle{mnras}
\bibliography{BG_tri}

\end{document}